\documentclass[journal]{IEEEtran}
\usepackage{graphicx}

\begin{document}

\title{On the Second Order Statistics \\ of the Multihop Rayleigh Fading Channel}

\author{Zoran~Hadzi-Velkov,~\IEEEmembership{}Nikola Zlatanov,~\IEEEmembership{}and George K. Karagiannidis~\IEEEmembership{\vspace{-3mm}}
\thanks{Accepted for IEEE TCOM.}
\thanks{Z. Hadzi-Velkov and N. Zlatanov are with the Faculty of Electrical Engineering and Information Technologies, Ss. Cyril and Methodius University, Skopje, Email: zoranhv@feit.ukim.edu.mk,
nzlatanov@manu.edu.mk}
\thanks{G. K. Karagiannidis is with the Wireless Communications Systems Group (WCSG), Department of Electrical and Computer Engineering, Aristotle University of Thessaloniki, Thessaloniki, Email: geokarag@auth.gr}
}

\markboth{}{Shell \MakeLowercase{\textit{et al.}}: Bare Demo of
IEEEtran.cls for Journals} \maketitle

\begin{abstract}
Second order statistics provides  a dynamic representation of a
fading channel and plays an important role in the evaluation and
design of the wireless communication systems. In this paper, we
present a novel analytical framework for the evaluation of
important second order statistical parameters, as the level
crossing rate (LCR) and the average fade duration (AFD) of the
amplify-and-forward multihop Rayleigh fading channel. More
specif\-ically, motivated by the fact that this channel is a
cascaded one and can be modeled as the product of $N$ fading
amplitudes, we derive novel analytical expressions for the average
LCR and the AFD of the product of $N$ Rayleigh fading envelopes
(or of the recently so-called $N*$Rayleigh channel). Furthermore,
we derive simple and eff\-icient closed-form approximations to the
aforementioned parameters, using the multivariate Laplace
approximation theorem. It is shown that our general results reduce
to the corresponding ones of the specif\-ic dual-hop case, previously published. Numerical
and computer simulation examples verify the accuracy of the
presented mathematical analysis and show the tightness of the
proposed approximations.
\end{abstract}


\section{Introduction}

\PARstart{M}{ultihop} communications have recently emerged as a
viable option for providing broader and more eff\-icient coverage
both in traditional (e.g. bent pipe satellites) and modern (e.g.
ad-hoc, WLAN) communications networks. In such systems, contrary
to conventional wireless networks, several intermediate terminals
operate as relays between the source and the destination
\cite{1}-\cite{Patel}.

Multihop transmissions can be categorized as either
non-regenerative (amplify-and-forward, AF) or regenerative
(decode-and-forward, DF), depending on the relay functionality. In
DF systems, each relay decodes its received signal and then
re-transmits this decoded version. On the other hand, in AF
systems, the relays just amplify and re-transmit their received
signal. Furthermore, the AF systems can use either channel state
information (CSI)-assisted relays \cite{1} or f\-ixed-gain relays
\cite{2} (also known as blind or semi-blind relays \cite{9}). A
(CSI)-assisted relay uses instantaneous CSI of the channel between
the transmitting terminal and the receiving relay terminal to
adjust its gain, whereas a f\-ixed-gain relay just amplif\-ies its
received signal by a f\-ixed gain \cite{2}\cite{9}. Note, that
systems with f\-ixed-gain relays perform close to the systems with
(CSI)-assisted relays \cite{2}, while their easy deployment and
low complexity make them attractive from a practical point of
view.

Several works in the open literature have provided performance
analysis of AF and DF systems in terms of bit error rate (BER) and
outage probability under different assumptions for the amplif\-ier
gain \cite{1}-\cite{Patel}. Among them, only two works deal with
the dynamic, time-varying nature of the underlying fading channel
\cite{10a}, \cite{Patel}, despite the fact that it is necessary
for the system's design or rigorous testing
\cite{Patel}-\cite{10c}. In \cite{10a}, the level crossing rate
(LCR) and the average fade duration (AFD) of multihop DF
communication systems over generalized fading channels is studied,
both for noise-limited and interference-limited systems, while
Patel et. al in \cite{Patel} provide useful exact analytical
expressions for the AF channel's temporal statistical parameters,
such as the auto-correlation and the LCR. However, the approach
presented in \cite{Patel} is limited only to the dual-hop
f\-ixed-gain AF Rayleigh fading channel.

In this paper, we study the second order statistics of the
multihop f\-ixed-gain AF Rayleigh fading channel. More
specif\-ically, motivated by the fact that this channel is a
cascaded one and can be modelled as the product of $N$ fading
amplitudes, we derive a novel analytical framework for evaluation
of the average LCR and the AFD of the product of $N$ Rayleigh
fading envelopes. Since the presented exact expressions are
computationally attractive only for small values of $N$, we derive
simple and yet eff\-icient closed-form approximations using the
multivariate Laplace approximation theorem [22, Chapter IX.5],
\cite{18}, which can be eff\-iciently used to evaluate the
aforementioned second order statistical parameters. These
important theoretical results are then applied to investigate the
second order statistics of the multihop Rayleigh fading channel.
Numerical and computer simulations validate the accuracy of the
presented mathematical analysis and show the tightness of the
proposed approximations.

The remainder of the paper is organized as follows: In the next
section, the second order statistics analysis of the product of
$N$ Rayleigh fading amplitudes is presented, providing exact and
tight approximated expressions for the LCR and the AFD. In section
IV, these theoretical results are applied to study the LCR and the
AFD of the f\-ixed-gain multihop relay fading channel. Section V
includes numerical and simulations results which validate the
proposed mathematical analysis while some concluding remarks are
given in section VI.


\section{Level Crossing Rate and Average Fade Duration of The Product of $N$ Rayleigh Envelopes }

Let $\{{X_i(t)}\}_{i=1}^N$ be $N$ independent and not necessarily
identically distributed (i.n.i.d.) Rayleigh random processes, each
distributed according to \cite{11}, \cite{12},
\begin{equation}\label{1}
f_{X_i}(x)=\frac{2x}{\Omega_i} \exp \left(
-\frac{x^2}{\Omega_i}\right) , \qquad x \geq 0,
\end{equation}
in an arbitrary moment $t$, where $\Omega_i=E\{X_i^2(t)\}$ is the
mean power of the $i$-th random process ($1\leq i \leq N$), with
$E\{\cdot\}$ denoting expectation.

If $\{{X_i(t)}\}_{i=1}^N$ represent received signal envelopes in
an isotropic scattering radio channel exposed to the Doppler
effect, they must be considered as time-correlated random
processes with some resulting Doppler spectrum. This Doppler
spectrum differs depending on whether a f\-ixed-to-mobile channel
\cite{11}-\cite{12} or mobile-to-mobile channel
\cite{13}-\cite{14} appears in the particular wireless
communication system. However, in both cases, it was found that
the time derivative of $i$-th envelope is independent from the
envelope itself, and follows the Gaussian probability distribution
function (PDF) \cite{11}-\cite{14}
\begin{equation}\label{2}
f_{\dot X_i}(\dot x )=\frac{1}{\sqrt{2\pi}\sigma_{\dot X_i }}
\exp\Big(-\frac{\dot x^2 }{2\sigma_{\dot X_i }^2}\Big) ,
\end{equation}
with variance
\begin{equation}\label{3}
\sigma_{\dot X_i }^2=\pi^2\Omega_i f_i^2 \,.
\end{equation}

If envelope $X_i$ is formed on a f\-ixed-to-mobile channel, then
\begin{equation}\label{3a}
f_i=f_{mi} \;,
\end{equation}
where $f_{mi}$ is the maximum Doppler frequency
shift induced by the motion of the mobile station
\cite{11}-\cite{12}. If envelope $X_i$ is formed on a
mobile-to-mobile channel, then
\begin{equation}\label{3b}
f_i=\sqrt{f_{mi}^{'2}+f_{mi}^{''2}} \,,
\end{equation}
where $f_{mi}^{'}$ and $f_{mi}^{''}$ are the maximum Doppler
frequency shifts induced by the motion of both mobile stations
(i.e., the transmitting and the receiving stations, respectively)
\cite{14}. It is important to underline that the maximum Doppler
frequency in a f\-ixed-to-mobile channel is
\begin{equation}\label{3c}
f_{d \max}=f_{mi} \,,
\end{equation}
whereas the maximum Doppler frequency in a mobile-to-mobile
channel is
\begin{equation}\label{3d}
f_{d \max}=f_{mi}^{'}+f_{mi}^{''} \,.
\end{equation}
The above results are essential in deriving the second-order
statistical parameters of individual envelopes, as the average LCR
and the AFD \cite{11}, \cite{12}, \cite{14}.

Below, we derive exact analytical and approximate solutions for
both of the above parameters for the product of the $N$ Rayleigh
envelopes,
\begin{equation}\label{4}
Y(t)=\prod_{i=1}^N X_i(t) \,,
\end{equation}
where $Y(t)$ in (\ref{4}) is $N*$Rayleigh random process or, at
any given moment $t$, $N*$Rayleigh random variable, following the
def\-inition given in \cite{15}.

For specif\-ied values $\{X_i\}_{i=1}^N = \{x_i\}_{i=1}^N$, the
product $Y$ is f\-ixed to the specif\-ic value
\begin{equation}\label{5}
y=\prod_{i=1}^N x_i \,.
\end{equation}

The LCR of $Y$ at threshold $y$ is def\-ined as the rate at which
the random process crosses level $y$ in the negative direction
\cite{11}. To extract LCR, we need to determine the joint PDF
between $Y$ and $\dot Y$, $f_{Y\dot Y}(y,\dot y)$, and to apply
the Rice's formula [17, Eq. (2.106)],
\begin{equation}\label{6}
N_Y(y)=\int_0^\infty \dot y f_{Y\dot Y} (y,\dot y) d\dot y \,.
\end{equation}

Our method does not require explicit determination of $f_{Y\dot
Y}(y,\dot y)$ in order to obtain analytically the LCR of the
$N*$Rayleigh random process, as presented below.

F\-irst, we need to f\-ind the time derivative of (\ref{4}), which
is
\begin{equation}\label{7}
\dot Y=Y\sum_{i=1}^N\frac{\dot X_i}{X_i} \,.
\end{equation}
Conditioning on the f\-irst $N-1$ envelopes $\{X_i\}_{i=1}^{N-1} =
\{x_i\}_{i=1}^{N-1}$, we have the conditional joint PDF $Y$ and
$\dot Y$ written as $f_{Y\dot Y|X_1\cdot\cdot\cdot X_{N-1}}(y,\dot
y|x_1,...,x_{N-1})$. This conditional joint PDF can be averaged
with respect to the joint PDF of the $N-1$ envelopes
$\{X_i\}_{i=1}^{N-1}$ to produce the required joint PDF of $Y$ and
$\dot Y$ as
\begin{eqnarray}\label{8}
f_{Y\dot Y}(y,\dot y) \qquad \qquad \qquad \qquad \qquad \qquad \qquad \qquad \qquad \quad \nonumber \\
=\int_{x_1=0}^\infty \cdots \int_{x_{N-1}=0}^\infty f_{Y\dot Y | X_1\cdots X_{N-1}}(y,\dot y | x_1,..., x_{N-1})\nonumber \\
\times \, f_{X_1\cdots X_{N-1}}(x_1,...,x_{N-1})dx_1\cdots dx_{N-1} \;\; \nonumber \\
=\int_{x_1=0}^\infty \cdots \int_{x_{N-1}=0}^\infty f_{Y\dot Y |
X_1\cdots X_{N-1}}(y,\dot y | x_1,..., x_{N-1})\nonumber \\
\times \, f_{X_1}(x_1)\cdots f_{X_{N-1}}( x_{N-1})dx_1\cdots
dx_{N-1} \,,
\end{eqnarray}
where to derive (\ref{8}) the mutual independence of the $N-1$
envelopes is used.

The conditional joint PDF $f_{Y\dot Y|X_1\cdot\cdot\cdot
X_{N-1}}(y,\dot y|x_1,...,$ $x_{N-1})$ can be further simplif\-ied
by fixing $Y = y$ and using the total probability theorem,
\begin{eqnarray}\label{9}
f_{Y\dot Y | X_1\cdots X_{N-1}}(y,\dot y | x_1,..., x_{N-1}) \qquad \qquad \qquad \qquad \quad \nonumber\\
=f_{\dot Y | Y X_1\cdots X_{N-1}}(\dot y |y, x_1,..., x_{N-1}) \qquad \qquad  \nonumber\\
\times \, f_{Y| X_1\cdots X_{N-1}}(y | x_1,..., x_{N-1}),
\end{eqnarray}
where each of the two multipliers in (\ref{9}) can be determined
from the above def\-ined individual PDFs and their parameters.

Based on (\ref{7}), the conditional PDF $f_{\dot Y|Y
X_1\cdot\cdot\cdot X_{N-1}}(\dot y|y,x_1,$ $...,x_{N-1})$ is
easily established to follow the Gaussian PDF with zero mean and
variance
\begin{eqnarray}\label{10}
\sigma_{\dot Y | Y X_1\cdots X_{N-1}}^2=\left(
y^2\sum_{i=1}^{N-1}\frac{\sigma_{\dot X_i}^2}{x_i^2}+ \sigma_{\dot
X_N}^2\prod_{i=1}^{N-1}x_i^2\right) \qquad \nonumber \\
=\sigma_{\dot X_N}^2 \left[1+y^2\left(\prod_{i=1}^{N-1}
\frac{1}{x_i^2}\right)\,\sum_{i=1}^{N-1}\frac{\sigma_{\dot
X_i}^2}{\sigma_{\dot X_N}^2}
\frac{1}{x_i^2}\right]\prod_{i=1}^{N-1}x_i^2 \,.
\end{eqnarray}

The conditional PDF of $Y$, given $\{X_i\}_{i=1}^{N-1} =
\{x_i\}_{i=1}^{N-1}$, that appears in (\ref{9}) is easily
determined in terms of the PDF of the remaining $N$-th envelope,
\begin{equation}\label{11}
f_{Y|X_1\cdots X_{N-1}}(y|x_1,..., x_{N-1}) =f_{X_N}\left(y
\prod_{i=1}^{N-1} \frac{1}{x_i}\right)
\prod_{i=1}^{N-1}\frac{1}{x_i} \,.
\end{equation}
Introducing (\ref{9}) and (\ref{11}) into (\ref{8}), then
(\ref{8}) into (\ref{6}), and changing the orders of the
integration, we obtain

\begin{eqnarray*}\label{12}
 N_Y(y) =  \int_{x_1=0}^\infty \cdots\int_{x_{N-1}=0}^\infty \left( \int_{\dot y=0}^\infty \dot
 y \right.
\qquad \qquad \qquad \qquad \qquad \nonumber\\
\times  \left. f_{\dot Y | Y X_1\cdots X_{N-1}}(\dot y |y,
x_1,..., x_{N-1})d\dot y \right)
\prod_{i=1}^{N-1}\frac{1}{x_i} \qquad \qquad \quad \,\, \nonumber \\
\times f_{X_N}\left (y\prod_{i=1}^{N-1}\frac{1}{x_i}\right )
f_{X_1}(x_1)\cdots f_{X_{N-1}}(x_{N-1})dx_1\cdots dx_{N-1}.
\end{eqnarray*}
\vspace{-6.0mm}
\begin{equation}
\end{equation}
The bracketed integral in (\ref{12}) is found using (\ref{10}) as
\begin{eqnarray}\label{13}
\int_{0}^\infty \dot y f_{\dot Y | Y X_1\cdots X_{N-1}}(\dot y |y,
x_1, \cdots , x_{N-1})d\dot y \qquad \qquad \quad  \nonumber\\
=\frac{\sigma_{\dot Y | Y X_1\cdots X_{N-1}}} {\sqrt{2\pi}} \qquad \qquad \qquad \qquad \qquad \qquad \qquad \quad \nonumber\\
=\frac{\sigma_{\dot X_N}} {\sqrt{2\pi}}
\left[1+y^2\left(\prod_{i=1}^{N-1}
\frac{1}{x_i^2}\right)\,\sum_{i=1}^{N-1}\frac{\sigma_{\dot
X_i}^2}{\sigma_{\dot X_N}^2}
\frac{1}{x_i^2}\right]^{1/2}\prod_{i=1}^{N-1}x_i \,.
\end{eqnarray}

By substituting (\ref{1}) and (\ref{13}) into (\ref{12}), we
obtain the f\-inal exact formula for the LCR as
\begin{eqnarray*}\label{14}
N_Y(y)= \frac{\sigma_{\dot X_N}}{\sqrt{2\pi}}\frac{2^Ny}{\Phi} \qquad \qquad \qquad \qquad \qquad \qquad \qquad \qquad \nonumber \\
\times \int_{x_1=0}^{\infty} \cdots\int_{x_{N-1}=0}^\infty\left[
1+y^2\left(\prod_{i=1}^{N-1}
\frac{1}{x_i^2}\right)\sum_{i=1}^{N-1}\frac{\sigma_{\dot
X_i}^2}{\sigma_{\dot X_N}^2} \frac{1}{x_i^2}\right]^{1/2} \nonumber \\
\times \exp\left[-\left(\frac{y^2}{\Omega_N}\prod_{i=1}^{N-1}
\frac{1}{x_i^2}+\sum_{i=1}^{N-1}\frac{x_i^2}{\Omega_i}\right)\right]dx_1\cdots
dx_{N-1} , \qquad
\end{eqnarray*}
\vspace{-4.0mm}
\begin{equation}
\end{equation}
where
\begin{equation}\label{15}
\Phi =\prod_{k=1}^{N}\Omega_k \,.
\end{equation}

In principle, (\ref{14}) together with (\ref{15}) provide an exact
analytical expression for the LCR of the product of the product of
$N$ Rayleigh envelopes (i.e., $N*$Rayleigh random process
\cite{15}). However, (\ref{14}) becomes computationally attractive
only for small values of $N$, such as $N=2$ and $N=3$, where it is
possible to apply multidimensional numerical integration (as
Gaussian-Hermite quadrature \cite{22}), included in most of the
well-known mathematical software packages.

The AFD of $Y$ at threshold $y$ is def\-ined as the average time
that the $N*$Rayleigh random process remains below level $y$ after
crossing that level in the downward direction,
\begin{equation}\label{16}
T_Y(y)=\frac{F_Y(y)}{N_Y(y)} ,
\end{equation}
where $F_Y(\cdot)$ denotes the cumulative distribution function
(CDF) of $Y$. Fortunately, $F_Y(\cdot)$ was derived recently in
closed-form [20, Eq. (7)], as
\begin{equation}\label{17}
F_Y(y)=G_{1,N+1}^{N,1} \left[\frac{y^2}{\Phi}\Bigg |
\begin{array}{cc}
 \qquad 1 \\
 \underbrace{1,1,\cdots ,1}_N, 0
\end{array}
 \right],
\end{equation}
where $G[\cdot]$ is the Meijer's $G$-function [21, Eq. (9.301)].
Note that Meijer's $G$-function is a standard built-in function in
well-known mathematical software packages, such as MAPLE and
MATHEMATICA.

\subsection{An Approximate Solution for the LCR}
Next, we present a tight closed-form approximation of (\ref{14})
using the multivariate Laplace approximation theorem [22, Chapter
IX.5], \cite{18} for the Laplace-type integral
\begin{equation}\label{18}
J(\lambda)=\int_{\textbf{x} \in D} u(\textbf{x}) \exp(-\lambda
h(\textbf{x}))d\textbf{x} ,
\end{equation}
where $u$ and $h$ are real-valued multivariate functions of
$\mathbf{x}=[x_1, \cdots , x_{N-1}]$, $\lambda$ is a real
parameter and $D$ is unbounded domain in the multidimensional
space $R ^{N-1}$.

A comparison of (\ref{14}) and (\ref{18}) yields
\begin{equation}\label{19}
u(\textbf{x})=\left[1+y^2\left(\prod_{i=1}^{N-1}\frac{1}{x_i^2}\right)\sum_{i=1}^{N-1}\frac{\sigma_{\dot
X_i}^2}{\sigma_{\dot X_N }^2}\frac{1}{x_i^2}\right]^{1/2} ,
\end{equation}

\begin{equation}\label{20}
h(\textbf{x})=\frac{y^2}{\Omega_N}\prod_{i=1}^{N-1}\frac{1}{x_i^2}+\sum_{i=1}^{N-1}\frac{x_i^2}{\Omega_i},
\end{equation}
and $\lambda=1$. A brief description of the multivariate Laplace
approximation theorem and its applicability conditions are
provided in the Appendix A.

Note, that in the case of (\ref{14}), all the applicability
conditions of the theorem are fulf\-illed. Namely, within the
domain of interest $D$, the function $h(\mathbf{x})$ has a single
interior critical point $\tilde \mathbf{x}=[\tilde x_1,\cdots
,\tilde x_{N-1}]$, where
\begin{equation}\label{21}
\tilde{x_i}=y^{1/N}\frac{\Omega_i^{1/2}}{\Phi^{1/(2N)}} ,\qquad
1\leq i\leq N-1 ,
\end{equation}
which is obtained from solving the set of equations $\partial h
/\partial x_i =0$, where $1\leq i\leq N-1$. The Hessian
$(N-1)\times (N-1)$ square matrix $\mathbf A$, def\-ined by (A.2),
is written as
\begin{equation}\label{22}
\mathbf{A}=\left[
\begin{array}{cccc}
8/\Omega_1 & 4/\sqrt{\Omega_1\Omega_2} & \cdots & 4/\sqrt{\Omega_1\Omega_{N-1}}\\
4/\sqrt{\Omega_2\Omega_1} & 8/\Omega_2 & \cdots & 4/\sqrt{\Omega_2\Omega_{N-1}}\\
. & . & \cdots & .\\
4/\sqrt{\Omega_{N-1}\Omega_1} & 4/\sqrt{\Omega_{N-1}\Omega_2} &
\cdots & 8/\Omega_{N-1}
\end{array}
\right ]
\end{equation}

By using induction, it is easy to determine that the $N-1$
eigenvalues of $\mathbf A$ are calculated as $\mu_i=4/\Omega_i$
for $1\leq i \leq N-2$, and $\mu_{N-1}=4N/\Omega_{N-1}$. Thus, all
eigenvalues of $\mathbf A$ are positive, which, by def\-inition,
means that the matrix $\mathbf A$ is positive def\-inite. By means
of the second derivative test, since the Hessian matrix $\mathbf
A$ is positive def\-inite at point $\tilde \mathbf x$, $h(\mathbf
x)$ attains a local minimum at this point (which in this case is
the absolute minimum in the entire domain $D$).

\noindent At this interior critical point $\tilde \mathbf {x}$, it
holds
\begin{equation}\label{23}
u(\tilde \mathbf {x})=\left (1+\sum_{i=1}^{N-1}\frac{\sigma_{\dot
X_i}^2}{\sigma_{\dot
X_N}^2}\frac{\Omega_N}{\Omega_i}\right)^{1/2}= \left (
1+\sum_{i=1}^{N-1}\frac{f_i^2}{f_N^2}\right)^{1/2} ,
\end{equation}
and
\begin{equation}\label{24}
h(\tilde \mathbf{x})=N\left (\frac{y^2}{\Phi}\right)^{1/N} \, ,
\end{equation}
where (\ref{23}) is obtained using (\ref{3}). Thus, by using
(A.3), it is possible to approximate (\ref{18}) for large
$\lambda$ as
\begin{eqnarray}\label{25}
J(\lambda) \approx \left(
\frac{2\pi}{\lambda}\right)^{(N-1)/2}\left[ \frac{1}{\det
(\mathbf{A})}\left(1+\sum_{i=1}^{N-1}\frac{f_i^2}{f_N^2}\right
)\right ]^{1/2} \nonumber \\
 \times \, \exp \left (-\lambda N\frac{y^{2/N}}{\Phi^{1/N}}\right) .
\end{eqnarray}
It is well-know that the determinant of the square matrix is equal
to the product of its eigenvalues, so $\det(\mathbf A)$ is
calculated as
\begin{equation}\label{26}
\det(\mathbf A)=\frac{N2^{2(N-1)}}{\prod_{k=1}^{N-1}\Omega_k}
=\frac{\Omega_N N2^{2(N-1)}}{\Phi} \,.
\end{equation}
Although approximation (\ref{25}) is proven for large $\lambda$
\cite{17}-\cite{18}, it is often applied when $\lambda$ is small
and is observed to be very accurate as well. Similarly to
\cite{19}, we apply the theorem for  $\lambda = 1$. Therefore, the
approximate closed-form solution for the LCR of $N*$Rayleigh
random process $Y$ at threshold $y$ is determined by
\begin{eqnarray*}\label{27}
N_Y(y) \approx \frac{\sigma_{\dot
X_{N}}}{\sqrt{2\pi}}\frac{2^Ny}{\Phi}J(1) = \frac{2y
(2\pi)^{N/2-1} \sigma_{\dot X_N}}{\Omega_N^{1/2} \Phi^{1/2}} \qquad \qquad \quad \nonumber \\
\times \, \left[\frac{1}{N} \left(
1+\sum_{i=1}^{N-1}\frac{f_i^2}{f_N^2}\right)\right]^{1/2}
\exp\left(-N\frac{y^{2/N}}{\Phi^{1/N}}\right) \quad \nonumber\\
= f_N\left[\frac{1}{N}\left (
1+\sum_{i=1}^{N}\frac{f_i^2}{f_N^2}\right)\right]^{1/2}\frac{(2\pi)^{N/2}y}{\Phi^{1/2}}
\exp\left(-N\frac{y^{2/N}}{\Phi^{1/N}}\right) \nonumber\\
= \left(\frac{1}{N}\sum_{i=1}^N
f_i^2\right)^{1/2}\frac{(2\pi)^{N/2}y}{\Phi^{1/2}}\,
\exp\left(-N\frac{y^{2/N}}{\Phi^{1/N}}\right) \,. \qquad \qquad
\end{eqnarray*}
\vspace{-7.0mm}
\begin{equation}
\end{equation}
The numerical results presented in Section IV validate the high
accuracy of the Laplace approximation applied for our particular
case.

Combining (\ref{17}) and (\ref{27}) into (\ref{16}), the AFD of
$N$ Rayleigh random process $Y$ at threshold $y$ is approximated
as
\begin{eqnarray}\label{27a}
T_Y(y) \approx  \left(\frac{1}{N}\sum_{i=1}^{N}f_i^2\right)^{-1/2}\frac{\Phi^{1/2}}{(2\pi)^{N/2}}\frac{1}{y} \qquad \qquad \qquad \quad \nonumber\\
\times \, G_{1,N+1}^{N,1} \left[\frac{y^2}{\Phi}\Bigg |
\begin{array}{cc}
\qquad 1 \\
\underbrace{1,1,\cdots ,1}_N, 0
\end{array}
 \right]\,\exp\left(N\frac{y^{2/N}}{\Phi^{1/N}}\right).
\end{eqnarray}

\subsection{Special Cases}
If the mean powers of all envelopes are assumed mutually equal,
$\{\Omega_i\}_{i=1}^{N}=\Omega$, (\ref{27}) reduces to
\begin{equation}\label{28}
N_Y(y)\approx\left(\frac{1}{N}\sum_{i=1}^N
f_i^2\right)^{1/2}\frac{(2\pi)^{N/2}y}{\Omega^{N/2}}\,\exp\left(-N\frac{y^{2/N}}{\Omega}\right),
\end{equation}
which is the approximate closed-form solution for the LCR of the
product of $N$ identically distributed Rayleigh envelopes.

Interestingly, {\it{when $N = 1$, (\ref{28}) further reduces to
the classic expression for the LCR of a Rayleigh-faded signal,
regardless of its mean power $\Omega$}}, i.e.
\begin{equation}\label{29}
N_Y(y)=f_1\sqrt{2\pi}\frac{y}{\sqrt{\Omega}}\,\exp\left(-\frac{y^2}{\Omega}\right)
,
\end{equation}
where $f_1=f_m$ for the f\-ixed-to-mobile channel
\cite{11}-\cite{12}, and $f_1=\sqrt{f_m^{'2}+f_{m}^{''2}}$ for the
mobile-to-mobile channel \cite{14}.

\section{Second Order Statiscs Of Multihop Transmissions}
Next, we apply the theoretical results of the previous
Section to analyze the second order statistics of the multihop
Rayleigh fading channel.
\subsection{System Model}
Let's consider a multihop wireless communications system,
operating over i.n.i.d. f\-lat fading channels (Fig. 1). The
source station $S$ communicates with the destination station $D$
through $N-1$ nodes $T_1$, $T_2$, ..., $T_{N-1}$, which act as
intermediate relays from one hop to the next. These intermediate
nodes are employed with non-regenerative relays with f\-ixed gain
$G_i$ given by
\begin{equation}\label{30}
G_i^2=\frac{1}{C_iW_{0,i}}
\end{equation}
with $G_0=1$ and $C_0=1$ for the source $S$. In (\ref{30}),
$W_{0,i}$ is the variance of the Additive White Gaussian Noise
(AWGN) at the output of the $i$-th relay, and $C_i$ is a constant
for the f\-ixed-gain $G_i$.

Assume that terminal $S$ is transmitting a signal $s(t)$ with an
average power normalized to unity. Then, the received signal at
the f\-irst intermediate node, $T_1$, at moment $t$, can be
written as
\begin{equation}\label{31}
r_1(t)=\alpha_1(t)s(t)+w_1(t) \,,
\end{equation}
where $\alpha_1(t)$ is the fading gain between $S$ and $T_1$, and
$w_1(t)$ is the AWGN at the input of $T_1$ with variance
$W_{0,1}$. The signal $r_1$ is then multiplied by the gain $G_1$
of the node $T_1$ and re-transmitted to node $T_2$, where its
received signal can be written as
\begin{equation}\label{32}
r_2(t)=G_1\alpha_2(t)\,(\alpha_1(t)s(t)+w_1(t))+w_2(t)
\end{equation}
with $\alpha_2(t)$ being the fading gain of the channel between
$T_1$ and $T_2$. Generally, the received signal at the $k$-th
relay $T_k$ ($k=1, 2,..., N-1$) is given by
\begin{equation}\label{33}
r_k(t)=G_{k-1}\alpha_k(t)r_{k-1}(t)+w_k(t) \,,
\end{equation}
f\-inally resulting in a total fading gain at the destination
station $D$, given by
\begin{equation}\label{34}
\alpha(t)=\prod_{i=1}^N\alpha_i(t)G_{i-1} \,.
\end{equation}

\begin{figure}
\centering
\includegraphics[width=3.5in]{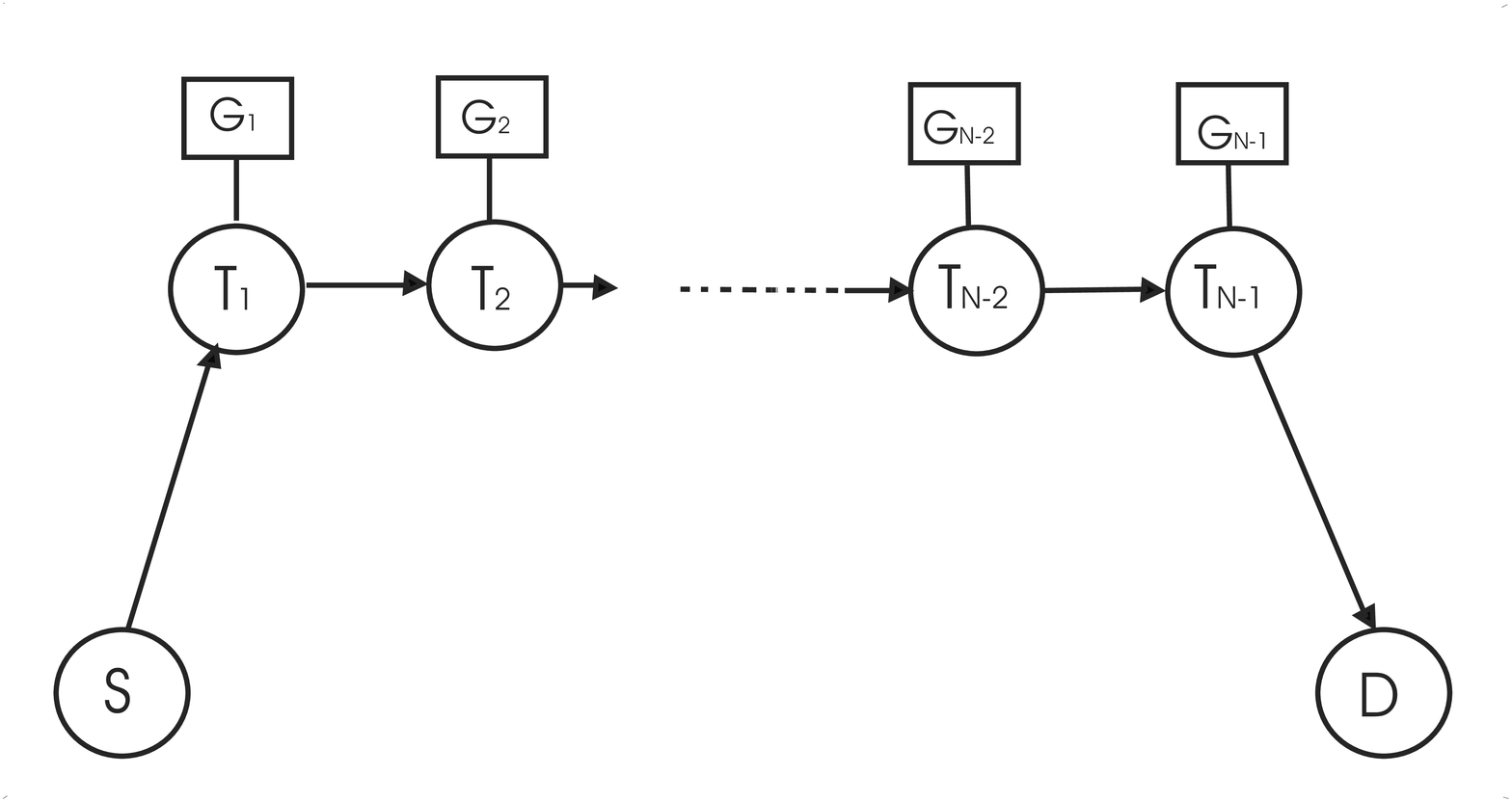}
\caption {Multihop wireless communication system} \label{fig_1}
\end{figure}

\subsection{LCR and AFD of Multihop Transmissions}
If the fading amplitude received at node $T_i$, $\alpha_i(t)$, is
a time-correlated (due to mobility of $T_{i-1}$ and/or $T_i$)
Rayleigh random process,  distributed according to (\ref{1}) at
any given moment $t$, with mean power
\begin{equation}\label{34b}
\hat \Omega_i=E\{\alpha_i^2(t)\} \, ,
\end{equation}
then the $i$-th element of the product in (\ref{34})
\begin{equation}\label{34c}
X_i(t)=\alpha_i(t)G_{i-1}
\end{equation}
is again a time-correlated Rayleigh random process, distributed
according to (\ref{1}), with mean power
\begin{equation}\label{34d}
\Omega_i=\hat \Omega_i\, G_{i-1}^2 \,.
\end{equation}

Comparing (\ref{4}) and (\ref{34}), we realize that the total
fading amplitude at the destination station $D$ (i.e., the
received desired signal without the AWGN) is described as the
$N*$Rayleigh random process $Y(t)=\alpha(t)$, whose average LCR
and AFD are determined in the previous Section.

Based on the system model from Fig. 1, if all stations are assumed
mobile with maximum Doppler frequency shifts $f_{mS} , f_{mD}$ ,
$f_{mi} (1\leq i\leq N-1)$ for source $S$, destination $D$ and
relays, respectively, then for the $i$-th hop
\begin{equation}\label{35a}
f_i^2=f_{m(i-1)}^2+f_{mi}^2
\end{equation}
with $f_{m0}=f_{mS}$ and $f_{mN}=f_{mD}$, and
\begin{equation}\label{35}
\sum_{i=1}^N f_i^2=f_{mS}^2+2\sum_{i=1}^{N-1} f_{mi}^2+f_{mD}^2
\,.
\end{equation}

Using (\ref{14}), (\ref{16}) and (\ref{17}), it is now
straightforward to obtain the exact expressions for the average
LCR and the AFD of the multihop Rayleigh fading channel, as
follows \setlength{\arraycolsep}{0.0em}
\begin{eqnarray}\label{35b}
& & N_\alpha(\alpha)= \frac{2^{N-\frac12} \sqrt{\pi} \alpha}{\Phi} \nonumber\\
&\times &  \int_{x_1=0}^{\infty} \cdots\int_{x_{N-1}=0}^\infty\left[ \Omega_N
f_N^2 +\alpha^2\left(\prod_{i=1}^{N-1}
\frac{1}{x_i^2}\right)\sum_{i=1}^{N-1}\frac{\Omega_i f_i^2}{x_i^2}
\right]^{1/2}
 \nonumber \\
&\times & \, \exp\left[-\left(\frac{\alpha^2}{\Omega_N}\prod_{i=1}^{N-1}
\frac{1}{x_i^2}+\sum_{i=1}^{N-1}\frac{x_i^2}{\Omega_i}\right)\right]dx_1\cdots
dx_{N-1}
\end{eqnarray}
\setlength{\arraycolsep}{5pt}
and
\begin{equation}\label{35c}
T_\alpha(\alpha)=\frac{1}{N_\alpha(\alpha)} \, G_{1,N+1}^{N,1}
\left[\frac{\alpha^2}{\Phi}\Bigg |
\begin{array}{cc}
 \qquad 1 \\
 \underbrace{1,1,\cdots ,1}_N, 0
\end{array}
 \right]
 ,
\end{equation}
respectively, where $f_i^2$ is given by (\ref{35a}) and $\Phi$ is
given by (\ref{15}).

It must be noted here that, for $N=2$, (\ref{35b}) is transformed
into a single integral of the form
\begin{eqnarray}\label{36b}
N_\alpha(\alpha)=\frac{4
\sqrt{\pi}\alpha}{\sqrt{2}\Omega_1\Omega_2} \qquad \qquad \qquad \qquad \qquad \qquad \qquad \quad \nonumber\\
\times \int_0^\infty\sqrt{\Omega_2 f_2^2+\Omega_1 f_1^2
\frac{\alpha^2}{x^4}}\,\exp\left[
-\left(\frac{\alpha^2}{x^2\Omega_2}+\frac{x^2}{\Omega_1}\right)\right]dx,
\end{eqnarray}
which, after changing integration variable $x$ with new variable
$t$ according $x=\alpha/t$, reduces to the known result [13, Eq.
(17)].

By combining (\ref{35}) with (\ref{27}) and (\ref{35}) with
(\ref{27a}), we also obtain approximate solutions for the average
LCR and AFD of the multihop Rayleigh fading channel as
\begin{eqnarray}\label{36}
N_{\alpha}(\alpha) \approx
\left[\frac{1}{N}\left(f_{mS}^2+2\sum_{i=1}^{N-1}
f_{mi}^2+f_{mD}^2\right)\right]^{1/2} \qquad \nonumber\\
\times \frac{(2\pi)^{N/2}\alpha}{\Phi^{1/2}}
\exp\left(-N\frac{\alpha^{2/N}}{\Phi^{1/N}}\right)
\end{eqnarray}
and
\begin{eqnarray*}\label{36a}
T_{\alpha}(\alpha) \approx
\left[\frac{1}{N}\left(f_{mS}^2+2\sum_{i=1}^{N-1}
f_{mi}^2+f_{mD}^2\right)\right]^{-1/2} \qquad \qquad \quad \nonumber \\
\times \frac{\Phi^{1/2}}{(2\pi)^{N/2}}\frac{1}{\alpha} \,\,
G_{1,N+1}^{N,1} \left[\frac{\alpha^2}{\Phi}\Bigg |
\begin{array}{cc}
 \qquad 1 \\ \underbrace{1,1,\cdots ,1}_N, 0
\end{array}
 \right]  \exp\left(N\frac{\alpha^{2/N}}{\Phi^{1/N}}\right),
\end{eqnarray*}
\vspace{-4.0mm}
\begin{equation}
\end{equation}
respectively, where $\Phi$ is given by (\ref{15}).

We see that (\ref{36}) and (\ref{36a}) approximate the average LCR
and AFD of the total fading amplitude for arbitrary power of the
fading amplitudes $\hat \Omega_i$, arbitrary relay gains $G_i$ and
arbitrary maximal Doppler shifts for the nodes $f_{mi}$.

It must be noted here that, for $N=2$, (\ref{36}) is an
eff\-icient closed-form alternative to the corresponding one given
by [13, Eq. (17)] for the dual-hop case. Furthermore, as it will
be shown in the next section, the proposed approximation is highly
accurate.

\subsection{Special Cases}
If we assume that $i$) all stations are mobile and induce same
maximal Doppler shifts (i.e., $f_{mS}=f_{mi}=f_{mD}=f_m$), $ii$)
the fading amplitudes in all hops have equal powers (i.e., $\hat
\Omega_i=\hat \Omega$), then (\ref{36}) reduces to
\begin{equation}\label{37}
N_{\alpha}(\alpha)\approx\sqrt{2}f_m\frac{(2\pi)^{N/2}\alpha}{\Phi^{1/2}}
\, \exp\left(-N\frac{\alpha^{2/N}}{\Phi^{1/N}}\right) \,,
\end{equation}
where, according to (\ref{15}), $\Phi=\hat
\Omega^N\,\prod_{i=1}^{N-1}G_i^2$.

If we assume that $i$) the destination station is f\-ixed (i.e.,
$f_{mD}=0$) but all other stations are mobile inducing same
maximal Doppler shifts (i.e., $f_{mS}=f_{mi}=f_m$), and $ii$) the
fading amplitudes in all hops have equal powers (i.e., $\hat
\Omega_i=\hat \Omega$), then (\ref{36}) reduces to
\begin{eqnarray*}\label{38}
N_{\alpha}(\alpha)\approx
f_m\left(\frac{2N-1}{N}\right)^{1/2}\frac{(2\pi)^{N/2}\alpha}{\Phi^{1/2}}
\exp\left(-N\frac{\alpha^{2/N}}{\Phi^{1/N}}\right) \,,
\end{eqnarray*}
\vspace{-3.0mm}
\begin{equation}
\end{equation}
where again $\Phi=\hat \Omega^N\,\prod_{i=1}^{N-1}G_i^2$.

\section{Numerical Results and Discussion}
In this Section, we provide some illustrative examples for the
average LCR and AFD of the fading gain process of the received
desired signal at the destination of a multihop non-regenerative
relay transmission system from Fig. 1. The numeric examples
obtained from the derived approximate solutions are validated by
extensive Monte-Carlo simulations over the system model described
in Section III .

Based on the system model from Fig. 1, we considered a multihop
transmission system consisted of a source terminal $S$, 4
f\-ixed-gain relays, and a destination terminal $D$. The
destination $D$ has f\-ixed position, whereas the source and all
relays are mobile and induce same maximal Doppler shift $f_m$.

The f\-ixed-gain relays are assumed semi-blind with gains in
Rayleigh fading channel calculated according to [2, Eq. (15)] and
[11, Eq. (19)]
\begin{equation}\label{38a}
G_{i,sb}^2=\frac{1}{\hat
\Omega_i}\,\exp\left(\frac{1}{\bar\gamma_i}\right)\,\Gamma\left(0,\frac{1}{\bar\gamma_i}\right),
\end{equation}
where $\bar \gamma_i=\hat \Omega_i/W_{0,j}$ is the mean SNR on the
$i$-th hop, and $\Gamma(\cdot,\cdot)$ is the incomplete Gamma
function. Relay gain calculated according to (\ref{38a}) assures
mean power consumption equal to that of a CSI-assisted relay,
whose gain inverts the fading effect of the previous hop while
limiting the output power at moments with deep fading.

Depending on the stations' mobility, we used two different 2D
isotropic scattering models for the Rayleigh radio channel on each
hop of the multihop transmission system. For the f\-ixed-to-mobile
channel (hop), we used the classic Jakes channel model
\cite{11}-\cite{12}. For the mobile-to-mobile channel (hop), we
used the Akki and Habber's channel model \cite{13}-\cite{14}. The
Monte-Carlo simulations of the latter were realized by using the
sum-of-sinusoids method proposed in \cite{20}-\cite{21}.

\begin{figure}
\centering
\includegraphics[width=3.5in]{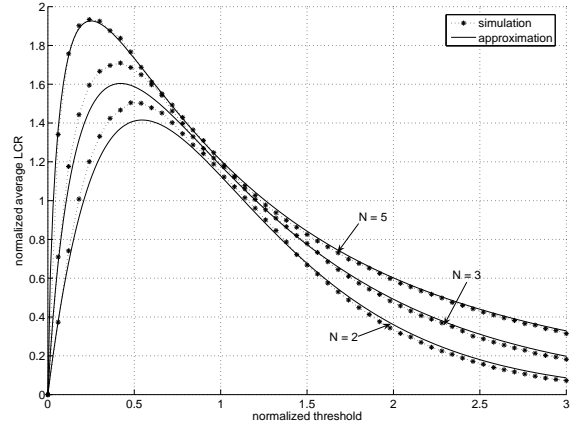}
\caption {Average level crossing rate, when $\hat \gamma_1=\hat
\gamma_2=\hat \gamma_3=\hat \gamma_4=\hat \gamma_5=\hat \gamma =
5$ dB} \label{fig_2}
\end{figure}

\begin{figure}
\centering
\includegraphics[width=3.5in]{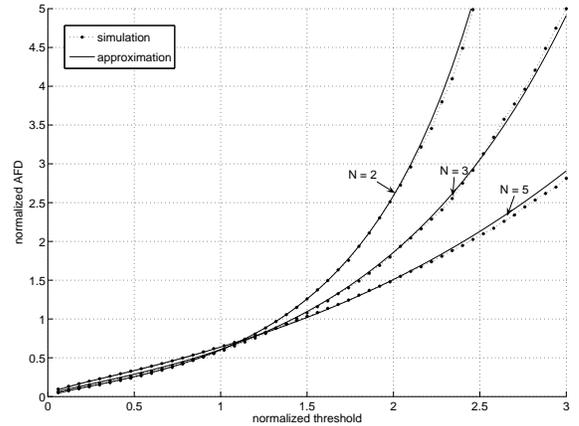}
\caption {Average fade duration, when $\hat \gamma_1=\hat
\gamma_2=\hat \gamma_3=\hat \gamma_4=\hat \gamma_5=\hat \gamma =
5$ dB} \label{fig_3}
\end{figure}

Each presented f\-igure depicts the received signal's normalized
LCR ($N_{\alpha}/f_m$) or normalized AFD ($T_{\alpha}f_m$) versus
the normalized threshold ($\alpha/\sqrt{\hat \Omega}$) at 3
different nodes along the multihop transmission system: at relay
$T_2$ (curve denoted by $N = 2$), at relay $T_3$ (curve denoted by
$N = 3$) and at the destination $D$ (curve denoted by $N = 5$).
Note that, when applying the considered scenarios in (\ref{36}) or
(\ref{38}), $\alpha$ and $\hat \Omega$ appear together as
$\alpha/\sqrt{\hat \Omega}$.

Figs. 2-5 assume equal power of the fading amplitudes in all hops
$ \hat \Omega_i=\hat \Omega$, and equal variance of the AWGN
$W_{0,i}=W_0$. Thus, $\bar\gamma_{i}=\bar\gamma$,
$G_{i,sb}=G_{sb}$ for $1 \leq i \leq 5$, so the mean of Rayleigh
random process $X_i(t)=\alpha_i(t)G_{i-1,sb}$ is calculated as
\begin{equation}\label{38b}
\Omega_i=\exp\left(\frac{1}{\bar\gamma}\right)\,\Gamma\left(0,\frac{1}{\bar\gamma}\right)=\Omega,
\quad 2\leq i \leq 5 ,
\end{equation}
whereas $\Omega_1=\hat \Omega$ is selected independently from the
AWGN, since $G_0=1$. In this case,
\begin{equation}\label{38c}
\Phi=\hat\Omega\,\exp\left(\frac{N-1}{\bar\gamma}\right)\,\left[\Gamma\left(0,\frac{1}{\bar\gamma}\right)\right]^{N-1}
\,.
\end{equation}

\begin{figure}
\centering
\includegraphics[width=3.5in]{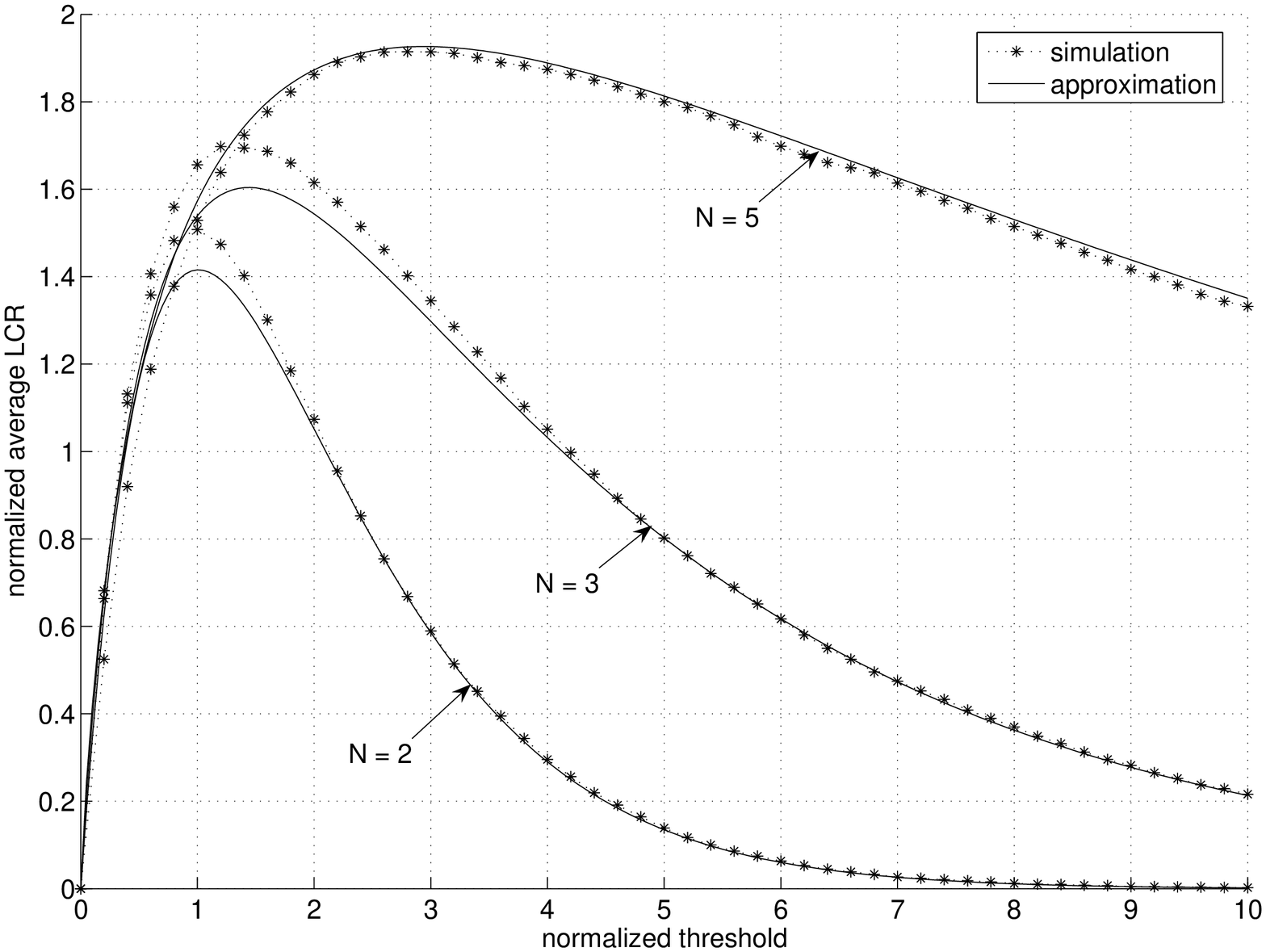}
\caption{Average level crossing rate, when $\hat \gamma_1=\hat
\gamma_2=\hat \gamma_3=\hat \gamma_4=\hat \gamma_5=\hat \gamma =
20$ dB} \label{fig_4}
\end{figure}

\begin{figure}
\centering
\includegraphics[width=3.5in]{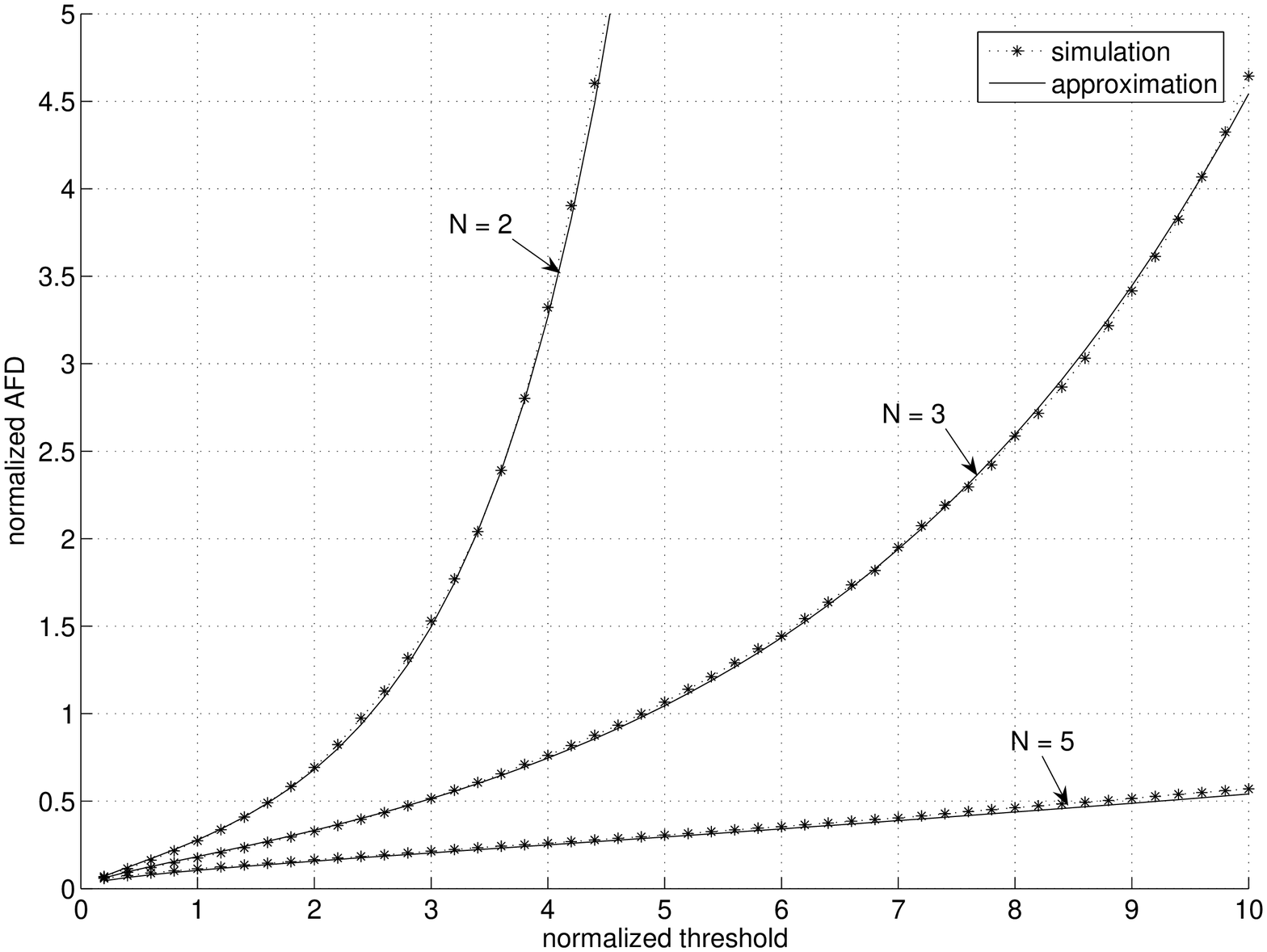}
\caption{Average fade duration, when $\hat \gamma_1=\hat
\gamma_2=\hat \gamma_3=\hat \gamma_4=\hat \gamma_5=\hat \gamma =
20$ dB} \label{fig_5}
\end{figure}


Figs. 6-7 assume equal powers of the fading amplitudes in all hops
$\hat\Omega_i=\hat \Omega$, and unequal variances of the AWGN
$W_{0,i}$. Thus, the mean of Rayleigh random process
$X_i(t)=\alpha_i(t)G_{i-1,sb}$ is calculated as
\begin{equation}\label{38c}
\Omega_i=\exp\left(\frac{1}{\bar\gamma_{i-1}}\right)\,\Gamma\left(0,\frac{1}{\bar\gamma_{i-1}}\right),
\quad 2\leq i\leq 5 ,
\end{equation}
whereas $\Omega_1=\hat \Omega$ is arbitrarily chosen. In this
case,
\begin{equation}\label{38c}
\Phi=\hat\Omega\,\exp\left(\sum_{i=1}^{N-1}\frac{1}{\bar\gamma_i}\right)\,\prod_{i=1}^{N-1}\Gamma\left(0,\frac{1}{\bar\gamma_i}\right)
\,.
\end{equation}

All comparative curves show an excellent match between the
approximate solution and the Monte-Carlo simulation for the
considered scenarios.

The LCR curves, presented in Figs. 2, 4 and 6, manifest a behavior
of a typical fading channel, since the average LCR increases with
the threshold until some maximum, and then decreases. The
maximized LCR and the maximizing threshold depend on number of
hops and per hop SNRs $\bar\gamma_{i}$. The AFD, presented in
Figs. 3, 5 and 7, also manifests a typical fading channel
behavior, since AFD continuously increases with the threshold.

Regardless on the per hop SNRs, for some specif\-ied threshold,
the $N*$Rayleigh fading signal typically remains less time in
fading with the increase of $N$ (Figs. 3, 5, and 7), whereas its
LCR increases (Figs. 2, 4, and 6). These differences are more
pronounced with the increase of the per hop SNRs. Similarly, for
some specif\-ied threshold, the LCR increases with the per hop
SNR, whereas its AFD decreases. Namely, for the considered range
of values for the $N$ and per hop SNR, $\Phi$ increases with
respect to both of those parameters, yielding to those
observations.


\begin{figure}
\centering
\includegraphics[width=3.5in]{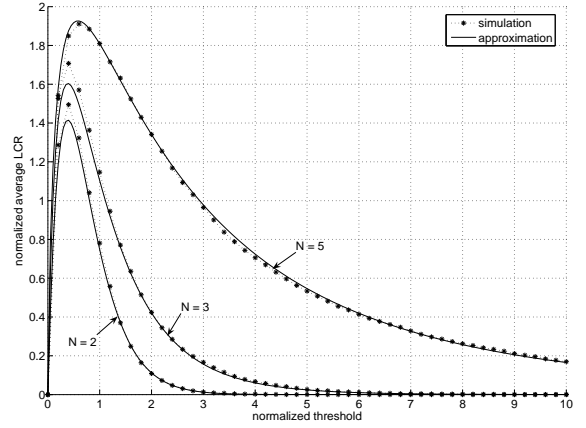}
\caption{Average level crossing rate, when $\hat \gamma_1 = 0$ dB, $\hat \gamma_2 = 10$
dB, $\hat \gamma_3 = 15$ dB, $\hat \gamma_4 = 15$ dB, $\hat
\gamma_5 = 20$ dB} \label{fig_6}
\end{figure}

\begin{figure}
\centering
\includegraphics[width=3.5in]{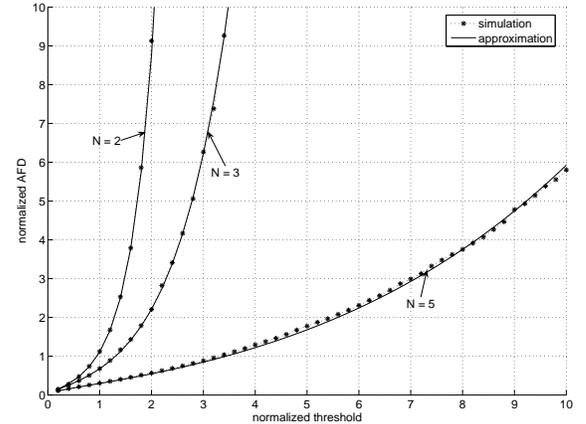}
\caption{Average fade duration, when $\hat \gamma_1 = 0$ dB, $\hat \gamma_2 = 10$ dB,
$\hat \gamma_3 = 15$ dB, $\hat \gamma_4 = 15$ dB, $\hat \gamma_5 =
20$ dB} \label{fig_7}
\end{figure}

\section{Conclusions}
In this paper, motivated by the fact that the multihop AF Rayleigh
fading channel is a cascaded one, we present a novel and general
analytical framework for the exact evaluation of important second
order statistical parameters, as the LCR and the AFD of this
channel. Moreover, we present simple and eff\-icient closed-form
approximations for the aforementioned parameters, using the
multivariate Laplace approximation theorem. The accuracy of the
presented mathematical analysis and the tightness of the proposed
closed-form approximations, were shown by numerical examples and
extensive Monte-Carlo simulations. The material presented in this
paper can be eff\-iciently used in determining the packet length
selection, power and bandwidth allocation over multiple hops, and
maximum delay/latency requirement \cite{10a}\cite{24}, or
determining the delay spread of frequency-selective multihop
channels \cite{25}. F\-inally, it could be useful in the study of
the second order statistics of the cooperative diversity systems.


\appendices
\section{Laplace Approximation Theorem}
Let $D$ be a possibly unbounded domain in the multidimensional
space $R^n$, $u$ and $h$ be real-valued multivariate functions of
$\mathbf{x}=[x_1,...,x_n]$, and $\lambda$ is a real parameter.
Consider the integral
\renewcommand{\theequation}{\thesection.\arabic{equation}}
\setcounter{equation}{0}
\begin{equation}\label{A1}
J(\lambda)=\int_{\textbf{x} \in D} u(\textbf{x}) \exp(-\lambda
h(\textbf{x}))d\textbf{x} .
\end{equation}
If $i$) the integral $J(\lambda)$ converges absolutely for all
$\lambda \leq \lambda_0$, $ii$) function $h$ has an absolute
minimum $\tilde{\mathbf{x}}=[\tilde x_1,\cdots ,\tilde x_n]$ at an
interior point of $D$ (this is turn implies that
$\tilde{\mathbf{x}}$ is a critical point of $h$, i.e., $\nabla
h(\tilde \mathbf{x} )=0$), and $iii$) the Hessian matrix
\begin{eqnarray}\label{A2}
\mathbf A&=&\left[\left(\frac{\partial^2h}{\partial x_i\partial
x_j}\right)\Bigg |_{\mathbf{x}=\mathbf{\tilde x}}\right]\nonumber\\
&=&\left[
\begin{array}{cccc}
\frac{\partial^2h(\mathbf{\tilde x})}{\partial x_1^2} & \frac{\partial^2h(\mathbf{\tilde x})}{\partial x_1 \partial x_2}& \cdots & \frac{\partial^2h(\mathbf{\tilde x})}{\partial x_1 \partial x_n}\\
\frac{\partial^2h(\mathbf{\tilde x})}{\partial x_2 \partial x_1} & \frac{\partial^2h(\mathbf{\tilde x})}{\partial x_2^2}& \cdots  & \frac{\partial^2h(\mathbf{\tilde x})}{\partial x_2 \partial x_n}\\
. & . & \cdots & .\\
\frac{\partial^2h(\mathbf{\tilde x})}{\partial x_n \partial x_1} & \frac{\partial^2h(\mathbf{\tilde x})}{\partial x_n \partial x_2}& \cdots  & \frac{\partial^2h(\mathbf{\tilde x})}{\partial x_n^2}\\
\end{array}
\right]
\end{eqnarray}
is positive def\-inite, then, for large $\lambda$,
\begin{equation}\label{A3}
J(\lambda)\approx
\left(\frac{2\pi}{\lambda}\right)^{n/2}\frac{u(\tilde
\mathbf{x})}{\sqrt{\det(\mathbf A)}}\exp(-\lambda h(\tilde
\mathbf{x})),
\end{equation}
where $\det(\cdot)$ represents the matrix determinant. The Laplace
approximation theorem was originally proven by Hsu \cite{18} for
$\lambda\rightarrow\infty$. It was observed in \cite{19} that in
many cases of interest the Laplace approximation performs very
well even in sub-asymptotic cases where $\lambda$ remains small.



\section*{Acknowledgement}
The authors wish to thank the Editor and the anonymous reviewers
for their valuable comments.



\end{document}